\documentstyle[epsfig]{mn}

\begin{document}


\title[CMB polarization and temperature anisotropies from
inflationary bubbles]{Effects of inflationary bubbles 
on the polarization and temperature anisotropies 
of the cosmic microwave background}
\author[C.Baccigalupi \& F.Perrotta]
{Carlo Baccigalupi \& Francesca Perrotta\\ 
SISSA/ISAS, Via Beirut 4 34014 Trieste, Italy\\
bacci@sissa.it\ ,\ perrotta@sissa.it}

\maketitle
 
\begin{abstract}
We predict the imprint of linear bubbly perturbations
on the polarization and temperature anisotropies  
of the cosmic microwave background (CMB). 

We analytically model a bubbly density perturbation
at the beginning of the radiation dominated era 
and we apply the linear theory of cosmological perturbations 
to compute its time evolution. At decoupling, it uniquely signs 
the CMB polarization and temperature anisotropy sky. 
As predicted by recent general work regarding 
spatially limited cosmological seeds, during evolution 
the perturbation propagates beyond the 
size of the bubble and reaches the CMB sound horizon at
the time considered. Therefore, its signal appears as a 
series of concentric rings, each characterized by its own 
amplitude and sign, on the scale of the sound horizon at 
decoupling ($\le 1^{o}$ on the sky). Polarization and 
temperature rings are strictly correlated; photons 
coming from the center of the bubble are not polarized, 
because of the spherical symmetry of the present problem. 
As expected for linear perturbations with size $L$ and density 
contrast $\delta$ at decoupling, $\delta T/T$ is roughly 
$\delta (L/H^{-1})^{2}$; the polarization is about $10\%$ of 
the temperature anisotropy. 

We predict the impact of a distribution of bubbles on the 
CMB polarization and temperature power spectra. 
Considering models containing both CDM Gaussian 
and bubbly non-Gaussian fluctuations, we simulate and 
analyze $10^{o}\times 10^{o}$ sky patches with angular 
resolution of about $3.5^{'}$. The CMB power associated with 
the bubbles is entirely on sub-degree angular scales 
($200\le l\le 1000$), that will be explored by the forthcoming 
high resolution CMB experiments with the percent precision. 
Depending on the parameters of the bubbly distribution we 
find extra-power with respect to the ordinary CDM Gaussian 
fluctuations; we infer simple analytical scalings of the 
power induced by bubbly perturbations and we constrain 
our parameters with the existing data. 
\end{abstract}

\section{Introduction}
\label{introduction}

World wide theoretical and experimental efforts 
to understand what we could learn from the Cosmic 
Microwave Background (CMB) are presently in progress. 
Anisotropies in the CMB should contain precious details 
about high energy physics processes
occurred in the early universe, still hidden to our knowledge. 
Reasonably, these processes left traces that have been stretched 
out to large and observable scales by an early era of 
accelerate expansion, called inflation; 
at decoupling between matter and radiation, these traces 
imprinted anisotropies in the CMB, thus being possibly detected 
and recognized. 

Recently, the treatment of the linear CMB inhomogeneities has been
put in a complete and organic form \cite{HSWZ}. 
At the same time, many experiments are at work to explore 
the CMB anisotropies toward smaller and smaller angular 
scales \cite{CMBPAST}; they shall culminate 
with the Planck mission of the next 
decade, that will provide the whole sky anisotropy map down 
to a minimum $\delta T/T\simeq 10^{-6}$ and an angular 
resolution of about $10^{'}$ \cite{PLANCK}.

According to the simplest inflationary phenomenology, at very 
high energy in the early universe ($T\simeq 10^{-5}$ Planck masses), 
a scalar field slowly rolls toward the minimum of its potential,
giving the non-zero vacuum energy responsible for the 
expansion itself. The quantum fluctuations generated during
this phase are stretched out by the inflationary expansion 
itself to large scales and become the
perturbations we observe today in the matter power spectrum
\cite{MFB}. However, this is not the only 
known inflationary mechanism. Many fundamental fields may act 
on stage and the effective potential may have several minima
separated by potential barriers. If this is the case 
quantum tunneling
occurs, and the nucleated bubbles are stretched out to large
scales just like the ordinary quantum fluctuations. At reheating the
energy stored in the bubble shell is converted into matter and 
radiation and a bubbly trace is left in the density distribution.
Tunneling processes are predicted in the context of first order 
inflation \cite{K}; the possibility that bubbles have left 
traces in the large scale matter distribution was considered 
in the last decade \cite{BUBBLE}, in connection with 
the increasing observational evidence 
of large voids in the galaxy distribution \cite{EPD}. 

The aim of this work is to accurately predict 
the CMB polarization and temperature anisotropies from 
bubbly density perturbations existing at decoupling, relics 
of the inflationary tunneling phenomena. 
In an early work \cite{BAO} we considered anisotropies 
from non-linear bubbles mainly for what concerns 
metric induced perturbations. Then, 
more recently \cite{B} we used an approximated method 
to compute the temperature anisotropies $\delta T/T$ from 
isolated inflationary bubbles in linear regime; we found that, 
just like a pebble in a pond, the bubbly perturbation generates 
anisotropy waves, traveling outward with the CMB sound velocity 
and reaching the scale of the sound horizon at the time 
we are examining it. Therefore, 
the image of an inflationary bubble at decoupling is a series of
concentric isothermal rings of different color (sign of $\delta T/T$)
on the scale of the sound horizon at decoupling ($\le 1^{o}$ 
in the sky). In a separate work \cite{ABOII} we explored the 
impact of a distribution of bubbles on the CMB temperature 
angular power spectrum, finding for a wide set of statistical 
parameters extra power on sub-degree angular scales.  

Here we compute the CMB polarization anisotropies induced
by inflationary bubbles, and we generalize previous 
results on the temperature in the context of the full linear 
theory of cosmological perturbations \cite{BKSMB}. 
We model the remnant of an inflationary bubble as a linear 
bubbly perturbation in the density distribution at the beginning 
of the radiation dominated era. Each Fourier mode is 
analytically calculated and numerically evolved following 
the linearized Einstein and Boltzmann equations. 
We use these results to study the signal from 
a single bubble. Regarding the size and density contrast 
of the bubbles, they are ultimately determined by the epoch 
of nucleation during inflation and 
by the properties of the tunneling barrier, 
so that these parameters depend on the details of the 
inflationary potential \cite{BUBBLE}; as in previous 
works \cite{B,ABOII} here we choose the 
bubble properties corresponding to 
the large voids observed in the galaxy distribution 
\cite{EPD}, that have deep cavities $\delta\rho /\rho\simeq 1$ 
and comoving radii of about $20\div 40h^{-1}$ Mpc; if they 
are of primordial origin, the evolution 
from decoupling to the present implies an overcomoving growth 
by a factor of 2, and the density contrast increases approximatively 
linearly with the scale factor \cite{OCCHIO}: thus, 
the voids corresponding to the present ones 
should have at decoupling a comoving radius 
$R=10\div 20h^{-1}$ Mpc and density contrast 
$\delta=10^{-3}\div 10^{-2}$, values that we adopt 
in this work. 

Starting from the signal from isolated bubbles, 
we predict the effect of a distribution of bubbles 
on the CMB temperature and polarization power spectra; 
as expected in scenarios admitting a bubble 
nucleating era during the inflationary slow roll 
\cite{BUBBLE}, we consider models containing a population 
of bubbles together with ordinary CDM Gaussian perturbations; 
in this context we simulate CMB sky maps and we compute 
the anisotropy angular power spectrum for various 
bubbly distribution parameters. As a general useful comment 
here, we wish to emphasize that both in the cases of a single 
bubble and a distribution of them, the signals treated 
here are definitely different from Gaussian perturbations. 
Gaussianity means that the fluctuations exhibit a completely 
disordered spatial distribution; only in this case the 
CMB angular power spectra tell us everything about the signal. 
On the contrary a bubbly perturbation is an ordered structure: 
it really possesses a spherical symmetry. Technically 
speaking, its signal is non-Gaussian; therefore, 
even if the CMB power spectrum is a powerful tool to investigate 
also non-Gaussian models, it does not fix completely the underlying 
fluctuations, and other methods should be adopted 
to investigate which type of non-Gaussian ordered pattern 
is present on the observed map. 
The general treatment of CMB anisotropies from cosmological structures 
having spatial finite extension and symmetries has been recently 
treated in detail \cite{PHASE}; we apply the general treatment 
developed in that work to the present case. 

The paper is organized as follows: in section II we recall the relevant
equations for our purposes and we expose the computational details; 
in section III we evolve the bubbly perturbation to study 
the corresponding CMB perturbation at different times and 
to predict the effect on the CMB polarization and temperature 
anisotropies from isolated bubbles; 
in section IV we consider several distributions 
of bubbles and we predict their imprint on the CMB angular 
power spectra, also giving simple scalings in terms 
of the distribution parameters; finally, section V
contains the conclusions.

\section{The equations system}
\label{theequations}

The evolution equations for fluid and CMB quantities may
be obtained by the Boltzmann and linearized Einstein equations
\cite{BKSMB,HSWZ}. A standard CDM flat background is assumed, 
including cold dark matter ($_{c}$), baryons ($_{b}$), 
photons ($_{\gamma}$) and three families of massless neutrinos 
($_{\nu}$). The background flat Friedmann Robertson Walker (FRW) 
metric is
\begin{equation}
\label{frw}
ds^{2}=a(\eta )^{2}\left(-d\eta^{2}+
dr^{2}+r^{2}d\Omega^{2}\right)\ ,
\end{equation}
where $\eta (t)=\int_{0}^{t} d\tau/a(\tau )$ is the 
conformal time. 
The background evolution is driven by the Einstein equation
\begin{equation}
{\dot{a}^{2}\over a^{2}}={8\pi G\over 3}a^{2}\sum_{x}\rho_{x}\ ,
\end{equation}
where the index $x$ runs over all the fluid species; 
the background parameters 
are $\Omega_{0}=1,h=.5,\Omega_{b}=0.05,
\Omega_{CDM}=1-\Omega_{b}$. 
The perturbed metric tensor is
\begin{equation}
\label{frwpert}
g_{\mu\nu}=a(\eta )^{2}(\gamma_{\mu\nu}+h_{\mu\nu})\ ,
\end{equation}
where $a(\eta )^{2}\gamma_{\mu\nu}$ represents the background. 
Since $h_{\mu\nu}\ll\gamma_{\mu\nu}$, a gauge freedom reduces 
the number of physically significant quantities in 
the perturbation metric tensor; 
in this work we adopt the generalized Newtonian gauge in which 
the two scalar perturbed metric component are $\Psi =h_{00}/2$ 
and $2\Phi =h_{11}=h_{22}=h_{33}$ \cite{BKSMB}; 
concerning the fluid, $\delta_{x}$ indicates the density 
contrast $\delta\rho /\rho$ for the species $x$ and 
$v_{x}$ its peculiar velocity in the background metric. 
Unless otherwise specified, all the equations for the 
perturbed quantities in the remaining part of this section 
are written in Fourier space, and the wavenumber argument 
is omitted. 

The equations for the matter species are: 
\begin{equation}
\label{c}
\dot{\delta}_{c}=-kv_{c}-3k^{2}\dot{\Phi}
\ ,\ \dot{v}_{c}=-{\dot{a}\over a}v_{c}+k\Psi\ ,
\end{equation}
\begin{equation}
\label{b}
\dot{\delta}_{b}=-kv_{b}-3k^{2}\dot{\Phi}
\ ,\ \dot{v}_{b}=-{\dot{a}\over a}v_{b}+k\Psi +
{4\rho_{\gamma}\over 3\rho_{b}}an_{e}\sigma_{T}
(v_{\gamma}-v_{b})\ ,
\end{equation}
where $\dot{\tau}=ax_{e}n_{e}\sigma_{T}$ is the differential
optical depth; $\sigma_{T}$ is the Thomson scattering 
cross section, $n_{e}$ is the electron 
number density and $x_{e}$ the ionization fraction \cite{HS}. 

Now we give the expressions for the CMB temperature 
and polarization perturbations in the context of a 
spherical perturbation field; these formulas have been 
recently developed \cite{PHASE}. As customary, we indicate 
the temperature perturbation $\delta T/T$ with $\Theta$ and 
the linear polarization amplitude with the Stokes 
parameters $Q$, $U$. 

Temperature perturbation carried by CMB photons scattered 
on an arbitrary direction $\hat{n}$ at a spacetime position 
$(\eta ,\vec{r})$ around a spherical structure is given by 
\begin{equation}
\label{thetasphere}
\Theta = \sum_{l}
P_{l}(\hat{n}\cdot\hat{r})
\int {k^{2}dk\over 2\pi^{2}}\Theta_{l}(\eta ,k)j_{l}(kr)\ ,
\end{equation}
where $P_{l}$ and $j_{l}$ are respectively Legendre polynomials 
and fractional order Bessel functions. All the specifications 
regarding the perturbation source are encoded in the 
$\Theta_{l}$ coefficients, which obey evolution 
equations that we shall write in a moment. 
Concerning the polarization, we have to specify the 
frame axes describing the polarization tensor on the 
plane perpendicular to $\hat{n}$. We choose them parallel 
and perpendicular to the plane formed by $\hat{n}$ and 
$\hat{r}$; with this choice, the polarization tensor 
is described only by $Q$ \cite{PHASE} in the following way:
\begin{equation}
\label{qsphere}
Q = \sum_{l\ge 2}
\sqrt{(l-2)!\over (l+2)!}
P_{l}^{2}(\hat{n}\cdot\hat{r}) 
\int{k^{2}dk\over 2\pi^{2}}
E_{l}(\eta ,k)j_{l}(kr)\ .
\end{equation} 
$P_{l}^{2}$ are second order Legendre polynomials; 
this is a significant difference from (\ref{thetasphere}): 
correctly, they make photons moving radially not 
polarized ($P_{l}^{2}(\pm 1)=0$). Again all the 
characteristics of the perturbation source are contained 
in the $E_{l}$ quantities. Now we give the evolution 
equations for the $E_{l}$ and $\Theta_{l}$ coefficients. 
They are the amplitudes of polarization 
and temperature CMB perturbations expanded in scalar and 
tensor spherical harmonics \cite{HSWZ}, and obey the following 
equations: 
\begin{equation}
\label{theta0}
\dot{\Theta}_{0}=-{k\over 3}\Theta_{1}-\dot{\Phi}\ ,
\end{equation}
\begin{equation}
\label{theta1} 
\dot{\Theta}_{1}=k\Theta_{0}-{2\over 5}k\Theta_{2}+
\dot{\tau}(v_{b}-\Theta_{1})+k\Psi\ ,
\end{equation}
\begin{equation}
\label{theta2}
\dot{\Theta}_{2} = {2\over 3}k\Theta_{1}-{3\over 7}k\Theta_{3}-
\dot{\tau}\left({9\over 10}\Theta_{2}-{\sqrt{6}\over 10}E_{2}\right)\ ,
\end{equation}
\begin{equation}
\label{e2} 
\dot{E}_{2}=-{\sqrt{5}\over 7}kE_{3}-
\dot{\tau}\left({1\over 10}\Theta_{2}+{2\over 5}E_{2}\right)\ ,
\end{equation}
and for $l\ge 3$
\begin{equation}
\label{dttlboltzmann}
\dot{\Theta}_{l}=k\left[{l\over 2l-1}\Theta_{l-1}-
{l+1\over 2l+3}\Theta_{l+1}\right]-
\dot{\tau}\Theta_{l}\ ,
\end{equation}
\begin{equation}
\dot{E}_{l} = k
\left[{\sqrt{l^{2}-4}\over 2l-1}E_{l-1}-
{\sqrt{(l+1)^{2}-4}\over 2l+3}E_{l+1}\right]-
\dot{\tau}E_{l}\ .
\label{elboltzmann}
\end{equation}
In Newtonian gauge the lowest multipoles
are linked to the photon fluid quantities by
$\delta_{\gamma}=4\Theta_{0}$, $v_{\gamma}=\Theta_{1}$ and
$\pi_{\gamma}=12\Theta_{2}/5$. Massless neutrinos
can be treated as photons without the polarization and
Thomson scattering terms. 
There remain the equations 
for the gravitational potentials:
\begin{eqnarray}
k^{2}\Phi &=& 4\pi Ga^{2}\left(\rho_{c}\delta_{c}+
\rho_{b}\delta_{b}+\rho_{\gamma}\delta_{\gamma}+
\rho_{\nu}\delta_{\nu}\right)+\nonumber\\
&+&4\pi Ga^{2}{3\over k}{\dot{a}\over a}\left(
\rho_{c}v_{c}+\rho_{b}v_{b}+{4\over 3}\rho_{\gamma}v_{\gamma}
{4\over 3}\rho_{\nu}v_{\nu}\right)\ ,
\label{phi}
\end{eqnarray}
\begin{equation}
\label{psi}
-k^{2}(\Psi +\Phi)={8\pi G\over 3}
\left(\rho_{\gamma}\pi_{\gamma}+\rho_{\nu}\pi_{\nu}\right)\ .
\end{equation}
As it is known \cite{BKSMB}, at early times the above system
can be solved by using the tight coupling approximation
between photons and baryons. The multipole equations
are expanded in powers of $k/\dot{\tau}\ll 1$. The only
zero order terms are $\Theta_{0}$ and $\Theta_{1}$ from
(\ref{b},\ref{theta0},\ref{theta1}), and obey the following equations:
\begin{equation}
\label{thetazero}
\dot{\Theta}_{0}=-{k\over 3}\Theta_{1}-\dot{\Phi}\ ,
\end{equation}
\begin{equation}
\label{thetaone}
{d\over d\eta}\left[\left(1+
{3\rho_{b}\over 4\rho_{\gamma}}\right)\Theta_{1}\right]=
k\Theta_{0}+k\left(1+{3\rho_{b}\over 4\rho_{\gamma}}\right)\Psi\ \ ,
\end{equation}
where $\Theta_{1}$ is assumed to coincide with $v_{b}$ to the 
lowest order. Increasing the order in $k/\dot{\tau}$
the higher order multipoles are given by
\begin{eqnarray}
\label{tc2}
\Theta_{2}={k\over\dot{\tau}}{8\over 9}\Theta_{1}\ &,&\ 
E_{2}=-{\sqrt{6}\over 4}\Theta_{2}\ ,\\
\label{tcl}
\Theta_{l}={k\over\dot{\tau}}{l\over 2l-1}\Theta_{l-1}\ &,&\ 
E_{l}={k\over\dot{\tau}}{\sqrt{l^{2}-4}\over 2l-1}E_{l-1}\ .
\end{eqnarray}
We integrate in time the system (\ref{thetazero},\ref{thetaone},
\ref{tc2},\ref{tcl}) until $k/\dot{\tau}=.1$ 
occurs, thereafter integrating the complete equations; 
of course, care is taken that the results do not depend at all 
on this choice. 

We take adiabatic initial conditions: at early times 
$\delta_{c}=\delta_{b}=3\delta_{\gamma}/4=3\delta_{\nu}/4$ 
(all the velocity are initially zero) and 
the second member in equation (\ref{phi}) at $\eta =0$ 
is proportional for each Fourier mode to the initial 
perturbation spectrum, that we define now. 
As we mentioned in the introduction, a well understood 
property of nucleated bubbles in field theory is 
their spatial growth; once nucleated in a de Sitter 
background like the inflationary one, a bubble is stretched 
out to super-horizon scales like the ordinary slow-rolling 
Gaussian fluctuations. Thus, although the 
coupling between the inflationary fields and the known 
sector of particle physics is unknown, a reasonable 
and simple assumption is that at the end of inflation 
bubbles simply convert their energy density distribution 
in matter and radiation, leaving a bubbly trace in the 
cosmic fluid \cite{BUBBLE}. We parameterize the latter with the 
following model, that we already adopted in an earlier work \cite{B}: 
$$
{\delta\rho\over\rho} = -\delta\ \ 
\left(x\le 1-{\sigma\over 2}\right)\ \ ,
$$
\begin{equation}
{\delta\rho\over\rho} = A+B\sin\left[{2\pi\over\sigma}
\left(x+\sigma -1\right)\right]
\ \ \left(1-{\sigma\over 2}\le x\le 1\right)\ ,
\label{dc}
\end{equation}
$$
{\delta\rho\over\rho} = C+D\sin\left[{2\pi\over\sigma}
\left(x+\sigma -1\right)\right]
\ \ \left(1\le x\le 1+{\sigma\over 2}\right)\ ;
$$
$x=r/R$ is an adimensional radial coordinate with $R$ 
being the comoving radius of the bubble, and $\sigma$ is 
the shell thickness. 
The Fourier transform $(\delta\rho /\rho )_{k}=
\int (\delta\rho /\rho)({\vec{r}})e^{i\vec{k}\cdot{\vec{r}}}d^{3}x=
\int (\delta\rho /\rho)(r)(\sin kr /kr)4\pi r^{2}dr$ 
of this perturbation may be 
easily calculated analytically. The spectrum is of course strongly 
scale dependent: all the power is essentially concentrated around 
the wavenumber corresponding to the bubble radius $k\simeq 2\pi /R$, 
while oscillations at higher $k$ describe the bubble shell. 
The constants are determined by the requests of continuity and
compensation:
\begin{eqnarray}
D &=& C=B-{\delta\over 2}=A+{\delta\over 2}=\nonumber\\
=&\delta &
{(2-\sigma )^{2}+8-6\sigma^{2}(4-\sigma )/\pi^{2}
\over
16+4\sigma^{2}-24\sigma^{3}/\pi^{2}}\ .
\label{abcd}
\end{eqnarray}
The physically most relevant parameters are $\delta$ and $R$: the
former sets the amplitude of the perturbation and the latter its
comoving size. The shell thickness does not change the substance of the
result, and from now on we fix $\sigma =.3\ $.
In order to make the following results more clear, $\delta$ 
is normalized with the density contrast taken in 
the center of the bubble at decoupling between matter and 
radiation. 

The very final thing to say is about the initial 
conditions in the CMB equations (\ref{theta0},\ref{theta1},
\ref{theta2},\ref{e2},\ref{dttlboltzmann},\ref{elboltzmann}); 
everything is initially zero except for the lowest 
multipole of the temperature perturbation \cite{BKSMB} 
\begin{equation}
\label{adi}
\Theta_{0}(0)=-2\Psi(0)\ ,
\end{equation}
thus being proportional, for each Fourier mode, to the initial 
perturbation spectrum. 

In the next section we integrate the equations defined 
here to see how the bubbly perturbation evolves in time and 
how it would appear in an high resolution CMB anisotropy 
observation. 

\section{Temperature and polarization effects from a single 
bubble}

Let's put the bubbly perturbation (\ref{dc})
as initial condition of the equations and let it
evolve. We are interested in the behavior of the CMB
perturbation, so we take some photos at different 
times. First, look at figure (1). It
represents the pure temperature perturbation induced by 
a bubble with radius $R=20h^{-1}$ Mpc at the indicated
times expressed in redshift; on the $x$-axis we can read
the comoving distance from the center, and the photon
propagation direction is orthogonal to the radial
direction, $\hat{n}\cdot\hat{r}=0$, as indicated.
Panel $(a)$ shows the initial condition, that remains
unchanged until the horizon crossing, that for the 
scale that we have considered occurs
nearly at equivalence. Panel $(b)$ shows a central deep 
oscillation that appears at the indicated redshift;
also, a positive crest arises at a distance corresponding
roughly to the radius of the bubble. At the time indicated
in panel $(c)$ an opposite oscillation occurs in the center, 
forming a negative crest, while the positive one is traveling
outward. Finally, in panel $(d)$ we can see the situation
just before decoupling, thus resembling the real CMB 
anisotropies that we shall show in the following.
A central negative spot is left, together with a well
visible wave, with negative and positive crests, that is 
traveling outward with the sound velocity and that will
be photographed by the CMB decoupling photons. 
Up to an accuracy of percent, the relevant contribution 
to the perturbation is given by 
the lowest multipoles $\Theta_{0}$ and $\Theta_{1}$, 
that we have used for the computations in 
previous works \cite{B,ABOII};  in each panel we have plotted 
the contributions from these two terms only
(thin dashed lines); the difference is not visible since 
they agree at the level of accuracy specified. 

Remarkably, the same phenomenology occurs for polarization.
In figure (2) we plot the polarization amplitude $Q$ defined 
in the previous section as a function of the
comoving radial distance $r$ and for propagation perpendicular to 
the radius. The initial condition (panel $(a)$)
is vanishing since the first multipole of the polarization
components is $l=2$ and it is coupled with the same
momentum of the pure temperature perturbations, 
that is zero at the beginning; physically, the reason is 
that polarization is coupled to acoustic oscillations 
that do not occur when all the modes are outside 
the effective horizon. At the 
horizon crossing, panel $(b)$, a negative oscillation occurs, 
followed by an opposite one (panel $(c)$); note that the light 
coming from the center is not polarized, simply because 
light propagating radially in a spherically symmetric 
problem has to be unpolarized for a pure geometric 
reason. Just before
decoupling, a polarization wave is well visible (panel $(d)$);
it is placed at the same distance from the center as the
$\Theta$ wave in figure (1); this is an expected
feature, due to the coupling between the polarization and pure 
temperature perturbations, manifest in the system 
(\ref{theta0},\ref{theta1},\ref{theta2},\ref{e2},
\ref{dttlboltzmann},\ref{elboltzmann}). 

The general phenomenology is therefore similar to the waves
in a pond when a pebble is thrown into it. This is a general 
behavior valid for any cosmological structure having a 
spatially limited extension \cite{PHASE}. 
We can physically understand this point by considering the equation 
for the $l=0$ multipole of the temperature perturbation (to the 
lowest order in $k/\dot{\tau}$). As it is known \cite{HSWZ}, 
from (\ref{thetazero},\ref{thetaone}) we get 
\begin{eqnarray}
\ddot{\Theta}_{0}&+&3{d\over d\eta}
\left({3\rho_{b}\over 4\rho_{\gamma}}\right)c_{s}^{2}
\dot{\Theta}_{0}+c_{s}^{2}k^{2}\Theta_{0}=\nonumber\\
=-\ddot{\Phi}&-&3c_{s}^{2}{d\over d\eta}
\left({3\rho_{b}\over 4\rho_{\gamma}}\right)\dot{\Phi}-
{k^{2}\over 3} \Psi\ ,
\label{wavelike}
\end{eqnarray}
where $c_{s}=1/\sqrt{3(1+3\rho_{b}/4\rho_{\gamma})}$
is the sound velocity of the photon-baryon fluid.
The right hand side contains the gravitational forcing
terms, while the left hand side is just 
a damped wave equation.
For a localized initial inhomogeneity, the 
corresponding CMB perturbation 
propagates beyond the initial size, generating 
waves traveling outward with the sound velocity,
and reaching the size of the sound horizon at the time we 
are examining it.

Let us consider now the numbers. The central density contrast at
decoupling $\delta$ has been factored in the $y$-axis labels
of the figures. Roughly, the amplitude of $\Theta$ 
follows the known estimates existing for linear 
perturbations, $\Theta\simeq\delta (R/H^{-1})^{2}$ 
\cite{P}, where the Hubble radius is calculated at decoupling. 
The polarization amplitude is roughly 10$\%$ of the temperature 
perturbation; this is also an expected result, 
since the lowest multipole 
of the polarization is coupled with the quadrupole 
momentum of the temperature perturbation ($\Theta_{2}$) that
in turn is a first order effect in $k/\dot{\tau}$ 
with respect to the quantities in (\ref{thetazero},\ref{thetaone}), 
as is evident from (\ref{tc2}); 
the same consideration holds for the ordinary Gaussian 
perturbations \cite{HSWZ}. 

The graphs we have shown until now were merely photos of the
bubbly CMB perturbation at different times. Now we want to
predict in detail the CMB anisotropies. Unlike Gaussian 
scale-invariant perturbations, our bubbly CMB signal is 
localized and vanishes beyond a sound horizon from its center.
Therefore, its position along the photons path is an 
adjunctive degree of freedom with respect to the ordinary 
perturbation models and it has to be specified.
Because of the spherical symmetry of the bubble, it can be simply
the distance $d$ between its center and the last scattering 
surface (LSS) peak; the latter is defined as follows. CMB 
photons have probability $P(\eta)=\dot{\tau}e^{-\tau}$ 
($\tau$ is the differential optical depth defined in the 
previous section) to be last scattered between $z$ and $z+dz$; 
the detailed form of $P(\eta )$ depends on the particular 
cosmological scenario of course, nevertheless it is in 
most cases a Gaussian centered around $z\simeq 1100$ 
with width $\Delta z\simeq 50$ \cite{HS}. We define 
the LSS peak as the point from which CMB photons were 
last scattered with highest probability (maximum $P(\eta )$)
along the direction corresponding to the bubble center 
in the sky. 
Once $d$ is specified, we can get the CMB anisotropies via 
simple line of sight integrations:
\begin{equation}
\label{dttani}
\Theta (now,here,\hat{n})=\int_{0}^{now}
(\Theta +\Psi )[\eta ,r(\eta ,\hat{n}),\hat{n}]P(\eta)d\eta\ ,
\end{equation}
\begin{equation}
\label{quani}
Q(now,here,\hat{n})=\int_{0}^{now}
Q[\eta ,r(\eta ,\hat{n}),\hat{n}]P(\eta)d\eta\ ,
\end{equation}
where the Sachs-Wolfe effect on photons climbing 
out of the potential hills generated by the bubble, 
simply given by $\Psi$ \cite{HSWZ}, has been added; 
$d$, $\eta$ and the propagation direction $\hat{n}$ 
fix the radius $r$ in which we have to compute the CMB 
quantities in the integrals: as it is easy to see 
\begin{eqnarray}
r(\eta ,\hat{n})&=&[(d+\eta_{0}-\eta_{LSS})^{2}+\nonumber\\
+(\eta_{0}-\eta )^{2}&-&2(d+\eta_{0}-\eta_{LSS})
(\eta_{0}-\eta)\hat{n}\cdot\hat{n}_{c}]^{1/2}\ ,
\label{los}
\end{eqnarray}
where $\hat{n}_{c}$ is the direction of the radiation
coming from the center of the bubble. In fact, since 
$\eta_{0}-\eta$ is just the causal distance covered by a photon
last scattered at $\eta$ and reaching us today, equation 
(\ref{los}) is obtained with simple trigonometry \cite{PHASE}. 
Also we can define naturally the useful angle $\theta$ by
\begin{equation}
\hat{n}\cdot\hat{n}_{c}=\cos{\theta}\ ;
\end{equation}
it is simply the angle between the photon propagation direction
$\hat{n}$ and the direction corresponding to photons coming from
the center of the bubble in the sky.

Now we are ready to see the results. Figure (3) 
shows the CMB polarization (top) and temperature (bottom)
anisotropies for a bubble at decoupling and with the size 
indicated; signals are plotted as a function of $\theta$.

Look at the top panels first. 
They show the polarization anisotropy $Q$ 
of the light coming from the bubble; we recall that it represents
the difference between the temperature fluctuations polarized
along the axes parallel and perpendicular to the plane formed by the 
radial and the photon propagation direction. At small $\theta$, 
$Q$ is vanishing as it must be, since the radiation coming
from the bubble center is propagating radially and thus
must be not polarized. The polarization wave that we have 
shown in figure (2) has been
photographed by the decoupling photons here: it is well
visible for both the bubble sizes chosen at an angular
position corresponding to the sound horizon at
decoupling. Note that the bubble is confined to small
angles, say at about $\theta\le R/2H_{0}^{-1}\simeq 10^{'}$ for
the radii treated here. Therefore, the
polarization wave {\it beyond} the bubble size is a unique sign 
of its presence. Also note that in each panel we have
plotted three curves. The solid one correspond to a bubble
exactly lying on the LSS peak ($d=0$), while the short and long 
dashed ones correspond to a bubble lying backward ($d=-R$) and forward 
($d=+R$) with respect to the LSS peak: these curves allow us to see
the dependence of the anisotropy on the relative disposition of the
perturbation source with respect to the LSS; of course, if $d$ was
greater than the sound horizon, no signal at all could be visible.

Look now at the bottom panels. It is as before, except for the
fact that here is plotted the temperature anisotropy. 
Note the central negative spot, 
at the same location of the bubble; again, at a sound horizon from
the center, the $\delta T/T$ wave propagating outward with the
photon-baryon fluid sound velocity has been photographed
by the last scattered photons. We have previously computed the
curves for $\delta T/T$ \cite{B} by considering
only the monopole and the dipole ($\Theta_{0}$ and $\Theta_{1}$)
contributions to the signal (thin continue line); 
here we obtain again the previous results 
by considering all the multipoles,
showing at the same time their good accuracy.

For what concerns the numbers, a comparison with 
figures (2) and (3) shows how
the CMB anisotropies are weaker than in the 
images of the perturbation at different times. 
Together with the partial cancelation of the 
opposite signs of $\Theta$ and $\Psi$, 
this is also because of the line of sight
integration across the LSS; the size of the bubbles is 
comparable with the LSS width, and due to the oscillating
behavior of the signal, the effect of the integral is
a certain weakening of the resulting signal. 

Now, anticipating issues of the next section, 
we realistically predict what we should observe
in the CMB sky if a bubble was really present at decoupling 
on the observed patch. In figure (4) we have simulated 
$2^{o}\times 2^{o}$ portions of microwave sky. 
First look at the bottom panel. In each point we have plotted
the $Q$ Stokes parameter of the CMB anisotropy. The signal
consists of two components: a Gaussian scale invariant 
adiabatic CDM spectrum \cite{BKSMB} 
and a bubble with comoving size $R=20h^{-1}$ Mpc lying exactly on 
the LSS ($d=0$), just like the solid line of figure (3). 
This case could be 
realistic in the sense that if during inflation more than one field 
were on the scene, and at least one undergoes quantum tunneling and 
bubble nucleation, at reheating we should expect both the kinds of
perturbation of the figure, bubbly and Gaussian; 
also, they simply add together if the perturbations are linear. 
The fascinating circular imprint of the bubble is clearly visible 
well over the CDM contribution; the central density contrast in 
this case is $\delta=10^{-2}$. Note that the bubble 
lies on the center of the figure, within $10^{'}$ from
the center. The circular polarization wave is well 
visible and it is {\it outside} the 
bubble, as we have already pointed out. If $\delta$ is reduced, 
we go to the case of the upper panel, in which the bubble is also
present, but with a smaller amplitude, $\delta =5\cdot 10^{-3}$.
In this unlucky case the detection of the bubbly signal
would require the use of more sophisticated image analysis tools 
than the human eye. 

Finally, figure (5) shows the temperature signal
instead of the polarization. Note as the circular rings are well
evident in the bottom panel, where $\delta$ is higher.
Again, the bubble lies in the very central region, corresponding
to the small central dark spot; the rings around are a feature of the 
CMB own physics and are coupled, for extension and 
oscillating behavior, with the polarization ones.
The central dark spot is absent in the polarization case
due to the symmetry constraints previously mentioned. 

The cross detection, both in temperature and 
polarization anisotropies, of the CMB rings caused by a localized
perturbation like a bubble would be a powerful check if signals 
like the present ones should be really detected in the future 
high resolution maps from MAP and Planck \cite{PLANCK}.

In the next section we consider the imprint of a population 
of bubbles on the CMB anisotropy power spectrum. 

\section{Temperature and polarization effects from a 
distribution of bubbles}

Let us consider now a distribution of bubbles. 
Our purpose is to check whether, and under which 
hypothesis, a distribution of bubbles is able to 
make a distinctive imprint on the CMB angular 
power spectrum: let us spend a few words 
to recall its definition. Given a sky signal 
$s(\theta ,\phi )$, the angular power 
spectrum is a set of coefficients $C_{l}$s 
simply defined in terms of the 
coefficients $a_{lm}$ of its expansion into 
spherical harmonics: 
\begin{equation}
\label{cl}
C_{l}={1\over 2l+1}\sum_{m=-l}^{m=l}|a_{lm}|^{2}\ .
\end{equation}
Since it depends on the squared modules of 
the $a_{lm}$s coefficients, it is sensitive on the 
anisotropy {\it power} on the angular scale $\alpha$
roughly corresponding to $180/l$ degrees \cite{P}. 

This work contains some improvements with respect to 
previous results treating the CMB power spectrum from 
a distribution of bubbles \cite{ABOII}: first both 
polarization and temperature signals are computed here, 
and second the signals from far bubbles, namely up to a 
sound horizon from the LSS, are considered instead of only 
the closest ones at a distance of about a radius from the LSS. 

We analyze $10^{o}\times 10^{o}$ zero mean CMB anisotropy 
maps with $3.5^{'}\times 3.5^{'}$ squared pixels. 
Bubbly signals are randomly distributed within the patch, 
which is projected on the sky via the Healpix 
pixelization \cite{HEAL}, setting the signal outside 
the patch to zero; successively, the signal is 
developed into spherical harmonics using Fast Fourier 
Transform methods in order to get the $C_{l}$ coefficients. 
We analyze both maps containing only the signals from 
bubbles and the latter together with the CDM Gaussian 
perturbations as in figures \ref{f4},\ref{f5}. 

First, in order to test our analysis method, we 
consider only CDM fluctuations. Starting from 
the theoretical CMB power spectrum, we simulate 
a patch as specified above and we extract from 
the latter the power spectrum itself, thus reconstructing 
the theoretical one. In figure \ref{f6} we plot 
the reconstructed and 
theoretical CDM power spectra for temperature (up) 
and polarization (down) anisotropies. 
The spectra are in quite good agreement 
(roughly $10\%$ accuracy) for a wide range of angular scales, 
corresponding to $l\ge 50$, while of course they disagree 
as approaching the dimension of the patch 
(roughly corresponding to $l\simeq 20$); 
the oscillations in the simulated spactra 
are due to the sample variance on the sky patch, 
but for our purpose here this method works well. 

Let us define now the distribution of the bubbles. 
Because of the anisotropy propagation treated in the 
previous section, the signal on the LSS contains 
contributions from bubbles centered at a distance that 
extends up to a sound horizon $r_{s}$ at decoupling. 
Indicating with $R$ the radius of the bubbles, and 
with $x$ the fraction of volume occupied, the number 
of bubbles that we should consider in our patch is 
roughly 
\begin{equation}
\label{n}
n=x\cdot 2r_{s}\cdot {3\over 4\pi R^{3}}\cdot 
(2H_{0}^{-1})^{2}d\Omega\ ,
\end{equation}
where $d\Omega$ is the solid angle subtended by the patch 
in the sky and $2H_{0}^{-1}$ is the comoving distance 
of the LSS from us. In our simulations we adopted $R=20h^{-1}$ Mpc, 
so that for $r_{s}\simeq 100h^{-1}$ Mpc a $x=10\%$ volume 
fraction is covered by roughly 600 bubbles in the specified 
patch. 

\begin{figure}
\vskip 1.in
\hskip -.5in
\epsfig{file=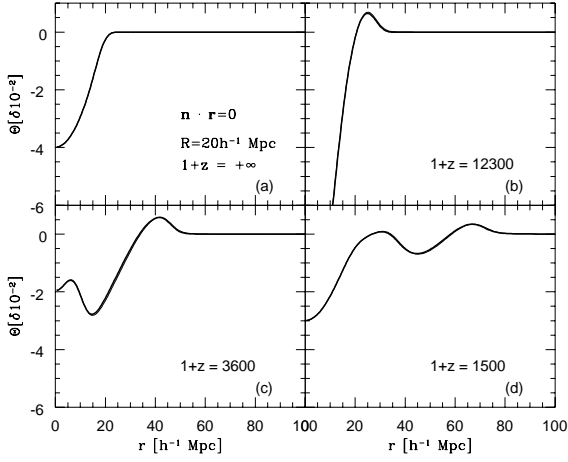,height=3.in,width=3.in}
\vskip -1.in
\caption{Radial behavior of the bubbly temperature perturbation
at different times. The initial condition remains unchanged until
the horizon reenter occurs. The successive oscillations generate
CMB perturbation propagating outward with the photon-baryon fluid
sound velocity.}
\label{f1}
\end{figure}

\begin{figure}
\vskip 1.in
\hskip -.5in
\epsfig{file=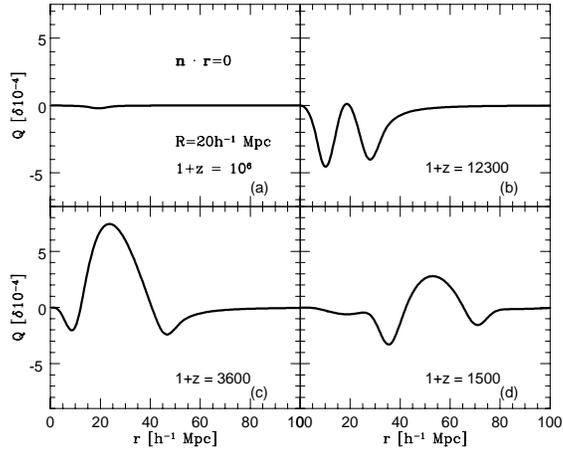,height=3.in,width=3.in}
\vskip -1.in
\caption{Radial behavior of the bubbly polarization perturbation
at different times. The initial condition remains unchanged until
the horizon reenter occurs. The successive oscillations generate
CMB polarization waves propagating outward as in the pure temperature
case. The substantial difference is the absence of central perturbation,
because of the spherical symmetry of the perturbations.}
\label{f2}
\end{figure}

\begin{figure}
\vskip 1.in
\hskip -.5in
\epsfig{file=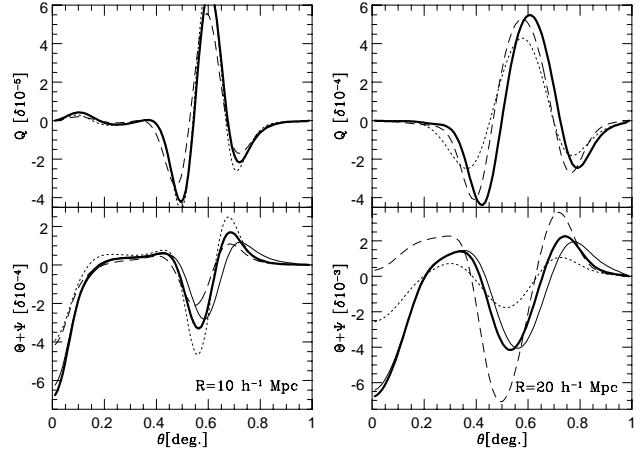,height=3.in,width=3.in}
\vskip -1.in
\caption{Polarization (up) and temperature (down) anisotropies 
from bubbles with the indicated radii intersecting 
exactly the last scattering surface. The sound waves have been 
photographed by the decoupling photons in both cases. 
For the polarization, the central perturbation spot is absent.
Thin long(short) dashed lines are anisotropies from bubbles placed
at a distance $+(-)R$ with respect to the last scattering surface. 
Thin continue lines in the bottom panels show the signals from 
the two first temperature multipoles $\Theta_{0}$ and $\Theta_{1}$.} 
\label{f3}
\vskip 5.in
\end{figure}

\begin{figure}
\vskip 1.in
\hskip -.25in
\epsfig{file=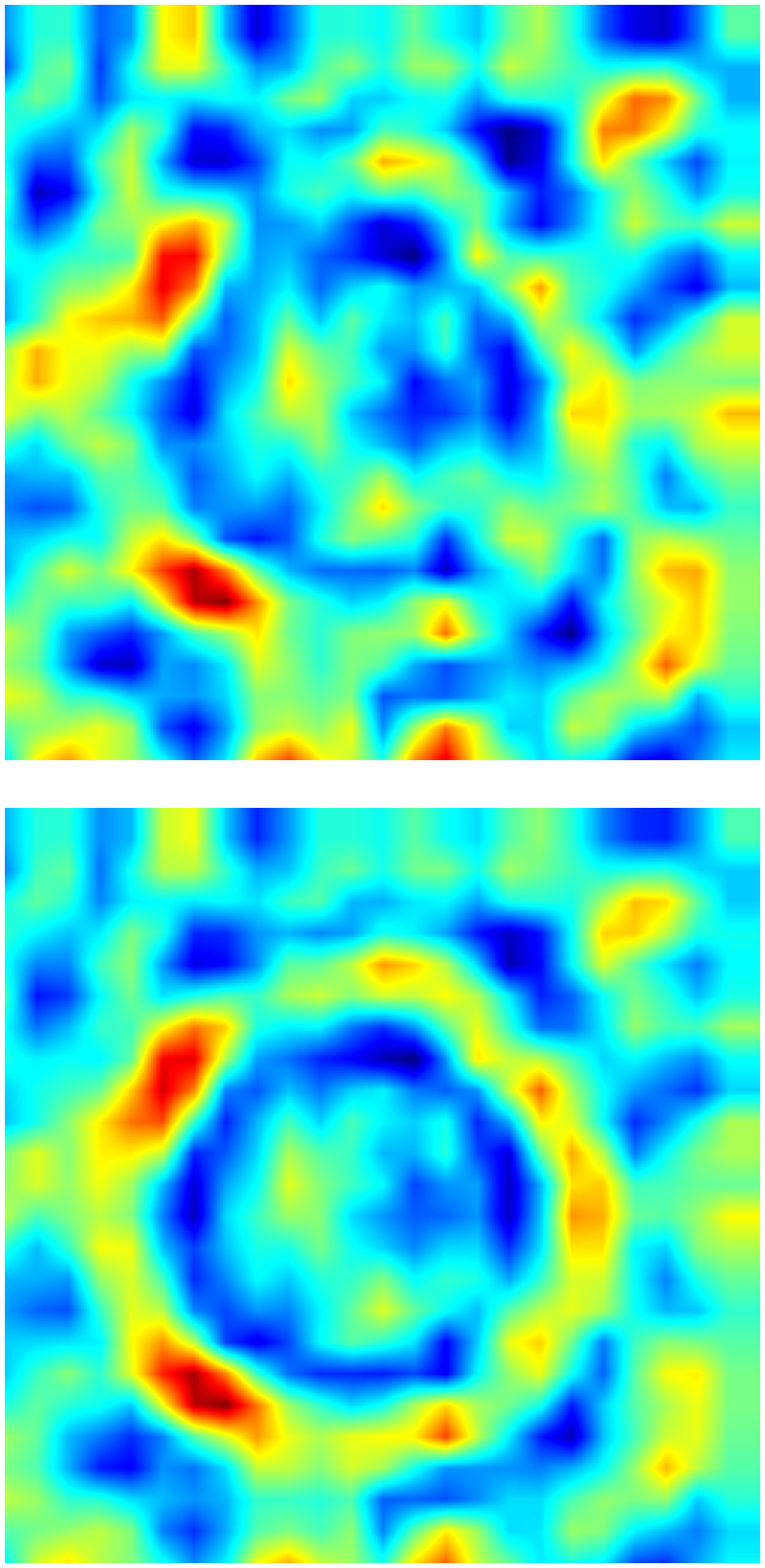,height=2.5in,width=3.5in}
\vskip 3.in
\caption{Realistic picture of the bubbly CMB polarization anisotropies.
The simulated patch is $2^{o}\times 2^{o}$. The bubble has radius
$R=20h^{-1}$
Mpc and central density contrast at decoupling 
$\delta =10^{-2}$ (bottom) and $\delta =5\times 10^{-3}$
(up). Together with the bubbly signal, the portion of sky simulated
contains a pure CDM Gaussian one.}
\label{f4}
\end{figure}

\begin{figure}
\vskip 1.in
\hskip -.25in
\epsfig{file=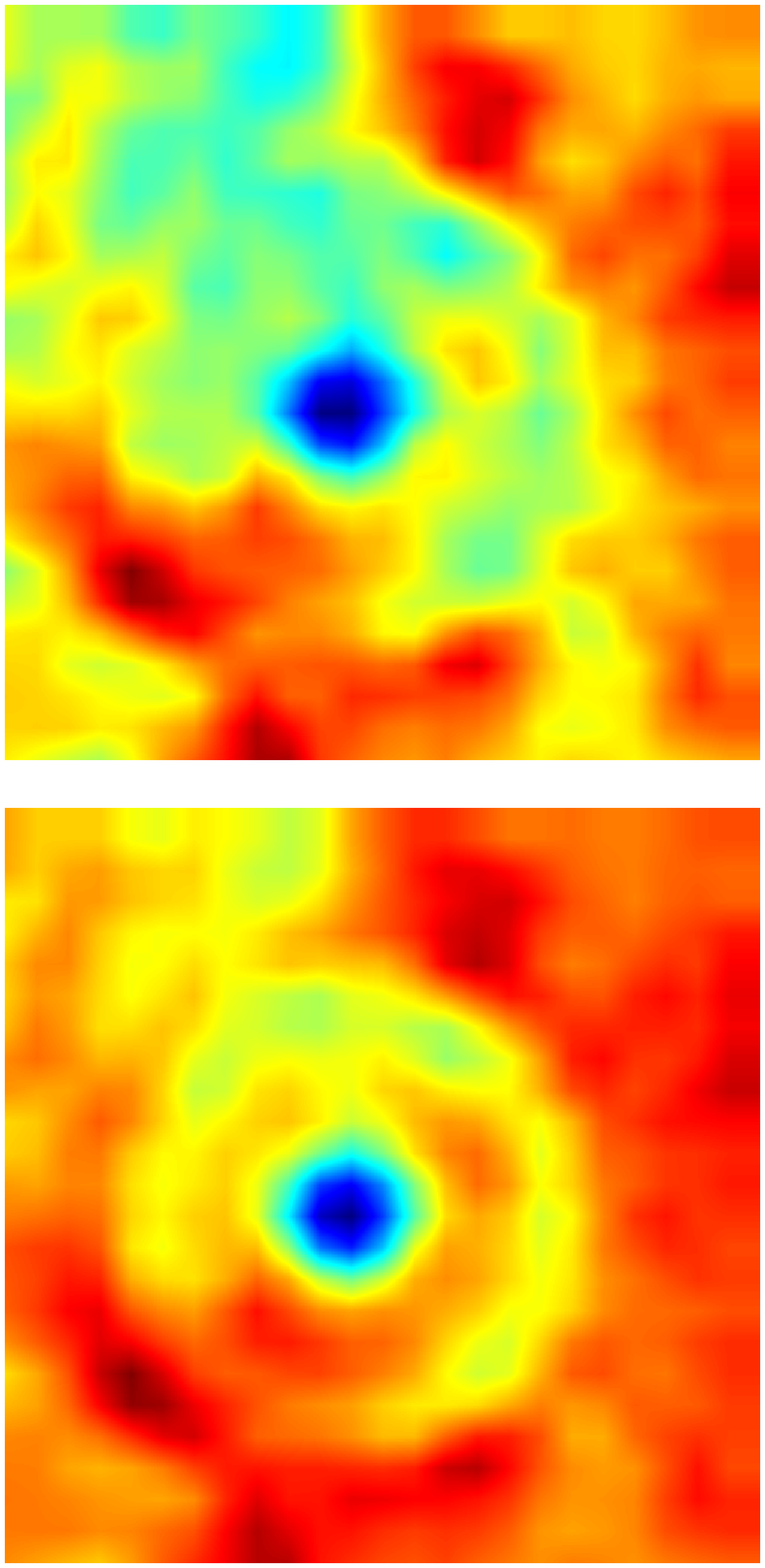,height=2.5in,width=3.5in}
\vskip 3.in
\caption{Realistic picture of the bubbly CMB temperature anisotropies.
The simulated patch is $2^{o}\times 2^{o}$. 
The bubble has radius $R=20h^{-1}$
Mpc and central density contrast at decoupling $\delta =10^{-2}$ 
(bottom) and $\delta =5\times 10^{-3}$ (up). Together with the bubbly 
signal, the simulated portion of the sky contains a pure CDM Gaussian 
one. Note the presence of the central dark spot, absent in the
polarization case.}
\label{f5}
\end{figure}

\begin{figure}
\vskip 1.in
\hskip -.25in
\epsfig{file=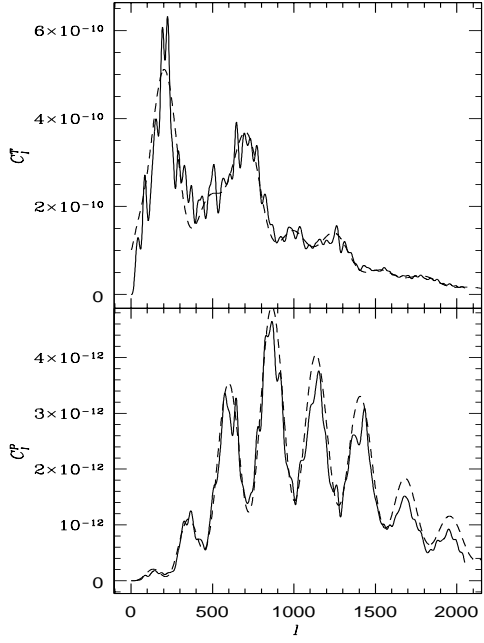,height=3.5in,width=4.5in}
\vskip 2.in
\caption{The theoretical (dashed line) CMB power 
spectra for the CDM Gaussian perturbations model 
adopted in this work has been simulated and reconstructed 
(solid line) in a $10^{o}\times 10^{o}$ sky patch. 
Temperature (up) and polarization (down) agree 
quite well for $l\ge 50$.}
\label{f6}
\end{figure}

\begin{figure}
\vskip 1.in
\hskip -.25in
\epsfig{file=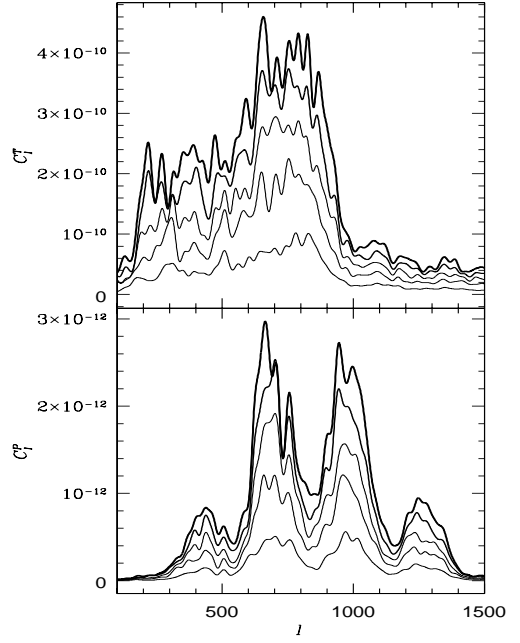,height=3.5in,width=4.5in}
\vskip 2.in
\caption{CMB temperature (up) and polarization (down) 
power spectra for bubbles with $R=20h^{-1}$ Mpc, 
$\delta =10^{-3}$ and 
filling volume fractions between $10\%$ (thin line) 
and $50\%$ (heavy line). Note how the signal is confined 
to sub-degree angular scales $l\ge 200$.}
\label{f7}
\end{figure}

First, let us analyze the effect of the bubbles only. 
Figure \ref{f7} shows the temperature (up) and polarization 
(down) spectra from bubbles filling a volume from $10\%$ 
(thin line) to $50\%$ (heavy line). The bubbles 
have central density contrast at decoupling 
$\delta\simeq 10^{-3}$. 
As expected for sub-horizon structures at decoupling, 
the whole signal lies on sub-degree 
angular scales, $l\ge 200$. In the temperature case, two 
region can be identified: the highest one describes 
the central hot spot of the bubbly signals and 
involves the corresponding multipoles at $l\simeq 800$; 
while the signals vanishes rapidly higher $l$s, 
it shows a sort of tail that reaches $l\simeq 200$, 
and that is generated by the anisotropy sound waves 
treated in the previous section extending the signal 
nearly up to one degree, see figures \ref{f1},\ref{f3},\ref{f5}. 
The spectrum form is completely different in the case of the 
polarization. This is an expected feature, since 
the polarization signal described in the previous section 
does not possess the central hot spot due to the spherical 
symmetry of the bubble. Instead, the polarization spectra 
contains peaks roughly at four angular scales that are related 
to the polarization waves departing from the center 
of the bubbles and extending up to about one degree and that 
are evident in figures \ref{f2},\ref{f3},\ref{f4}. 

\begin{figure}
\vskip 1.in
\hskip -.25in
\epsfig{file=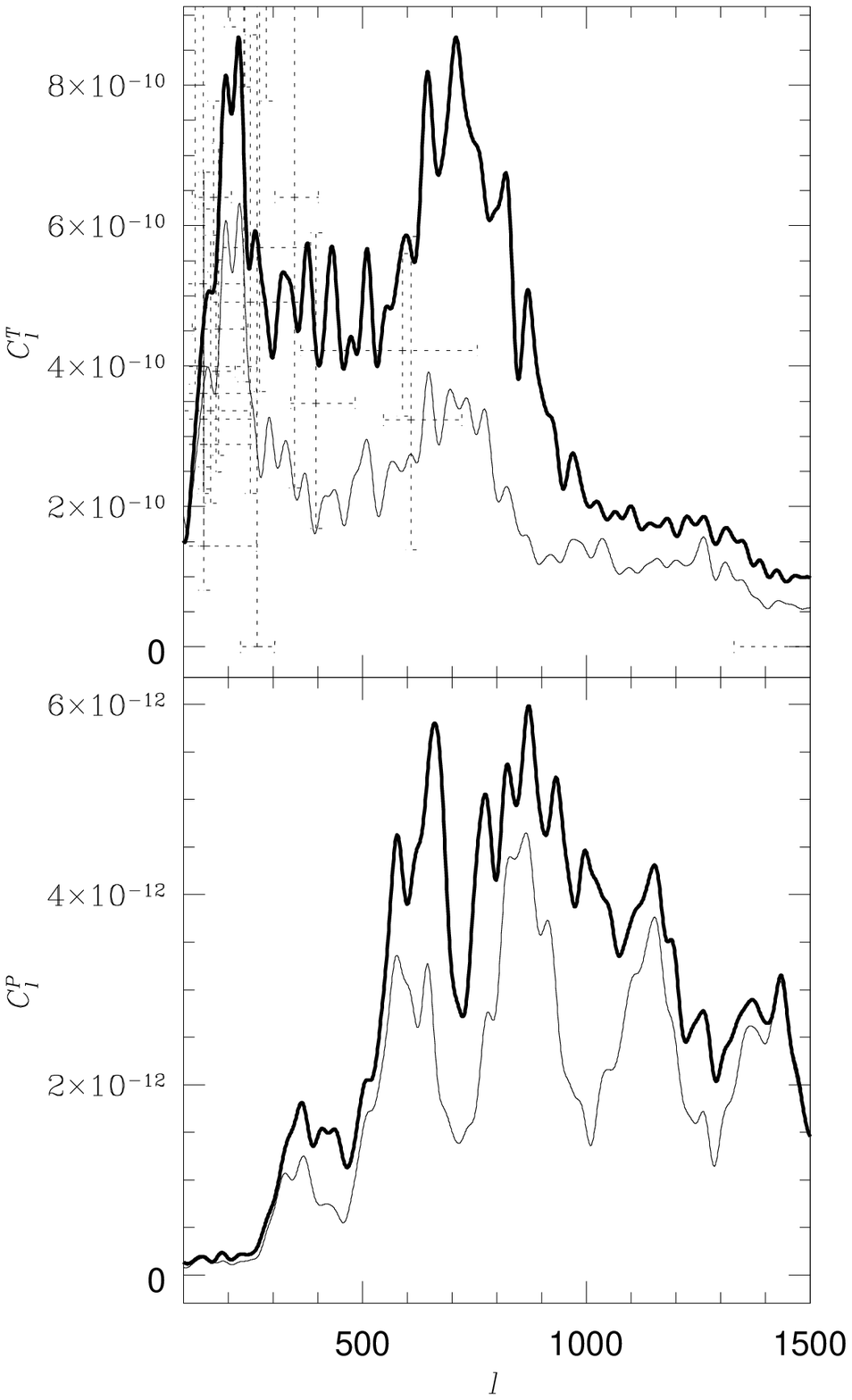,height=3.5in,width=4.5in}
\vskip 2.in
\caption{CMB temperature (up) and polarization (down) 
power spectra for Gaussian CDM perturbations and 
bubbles with $R=20h^{-1}$ Mpc, $\delta =10^{-3}$ and 
filling $50\%$ of the volume (heavy line). 
Also plotted are the purely Gaussian perturbations 
(thin line) and the existing experimental data on 
sub-degree angular scales $l\ge 200$.}
\label{f8}
\end{figure}

\begin{figure}
\vskip 1.in
\hskip -.25in
\epsfig{file=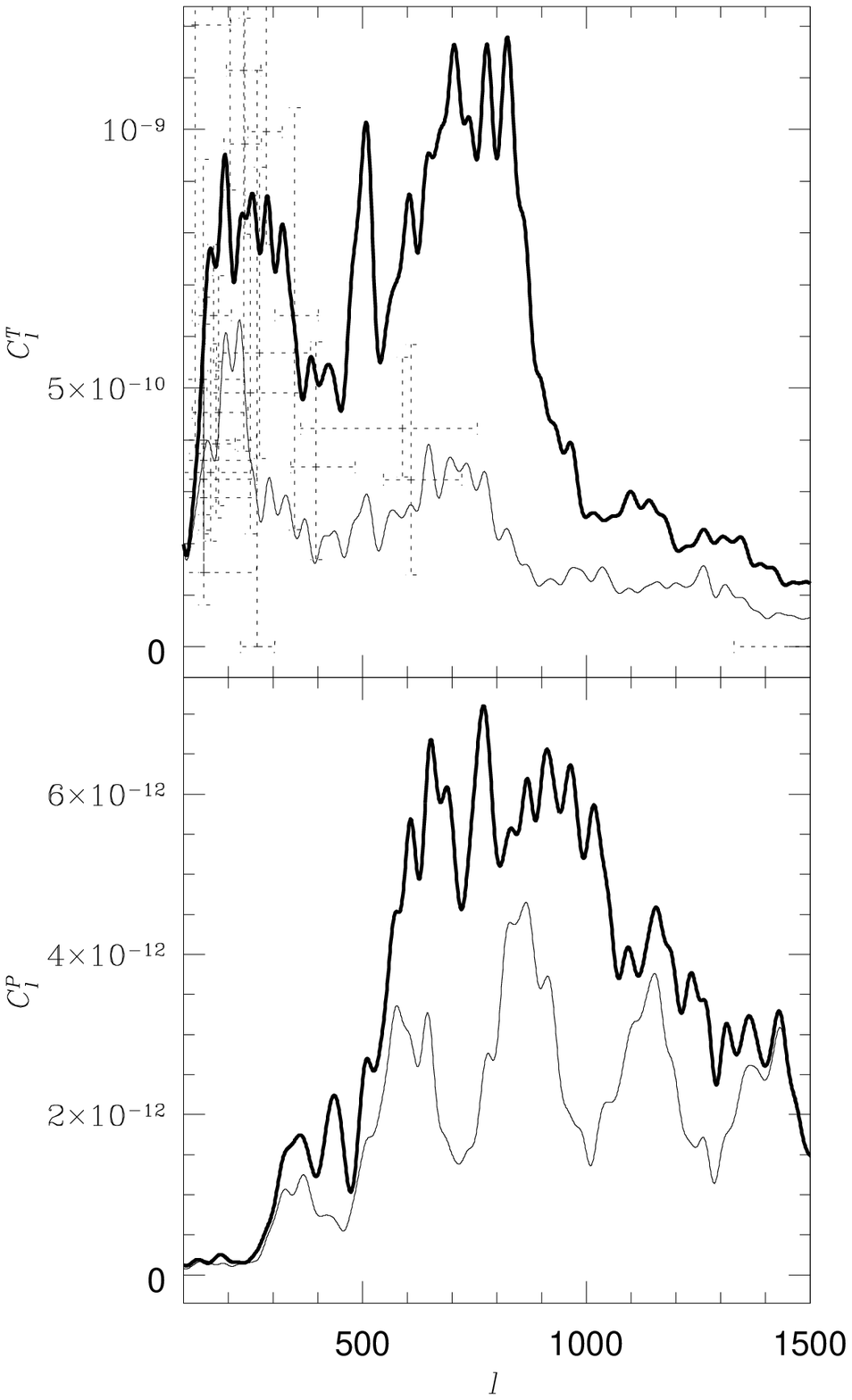,height=3.5in,width=4.5in}
\vskip 2.in
\caption{CMB temperature (up) and polarization (down) 
power spectra for Gaussian CDM perturbations and 
bubbles with $R=20h^{-1}$ Mpc,$\delta =3\cdot 10^{-3}$ 
and filling $10\%$ of the volume (heavy line). 
Also plotted are the purely Gaussian perturbations 
(thin line) and the existing experimental data on 
sub-degree angular scales $l\ge 200$.}
\label{f9}
\end{figure}

Let us come now to the evaluation of the scalings of the 
CMB power induced by bubbles. In this case all the bubbles 
have the same central density contrast $\delta$, so that 
their power spectra scale exactly as $\delta^{2}$. Moreover, 
as expected for spatially uncorrelated bubbles, the amplitude 
is quite linear into the occupied volume fraction $x$. 
In this work we do not show graphs relative to other radii, 
but since the bubbles are sub-horizon structures at 
decoupling \cite{P}, we checked that their signal scales 
roughly as $R^{2}$; in addition, as we mentioned in the previous 
section, the computation of the anisotropies causes an additional 
damping due to the line of sight integration across 
the LSS: the signal therefore drops by a factor 10 
passing from $20h^{-1}$ Mpc to $10h^{-1}$ Mpc, 
implying a scaling of about $R^{3.5}$. 
Putting together all these ingredients, and defining 
the bubbly spectral power as $C^{T,P}=l(l+1)C_{l}^{T,P}/2\pi$ 
calculated roughly at the maximum, we can infer the follow 
empirical scalings: 
\begin{eqnarray}
C^{T}&=&4\cdot 10^{-4}\delta^{2}
\left({x\over 50\%}\right)\cdot 
\left({R\over 20h^{-1}{\rm Mpc}}\right)^{7}\ ,
\label{powert}\\
C^{P}&=&3\cdot 10^{-6}\delta^{2}
\left({x\over 50\%}\right)\cdot 
\left({R\over 20h^{-1}{\rm Mpc}}\right)^{7}\ .
\label{powerp}
\end{eqnarray}
Note that these numbers, for what concerns temperature, 
lead to results that are bigger by nearly a factor of 2 than the 
scaling inferred previously \cite{ABOII}. This is 
probably because in our simulations more signal 
imprints the patch, since we considered the contributions 
from bubbles placed up to a CMB sound horizon away from 
the LSS. Since the recent measurements of CMB anisotropy 
power on sub-degree angular scales \cite{CMBPAST} 
indicate values roughly at the level 
$l(l+1)C_{l}/2\pi\simeq 4\cdot 10^{-10}$ 
in the region interesting for us, bubbles populations 
with $C^{T}$ larger then that values are likely to be 
excluded. This implies the following constraint on our 
parameters: 
\begin{equation}
\delta^{2}\left({x\over 50\%}\right)\left({R\over 
20h^{-1}{\rm Mpc}}\right)^{7}\le 10^{-6}\ .
\label{con}
\end{equation}
This condition is obtained considering only the power 
from bubbles, anyway it holds also when considering 
the power from Gaussian CDM perturbations, simply 
the two perturbations mechanisms are uncorrelated, 
so that their powers approximatively add. Indeed, 
figures \ref{f8} and \ref{f9} show the CMB angular 
power spectra in the cases in which bubbles and 
CDM Gaussian perturbations are present in 
the simulated patch. Also plotted are the existing data 
on $l\ge 200$ \cite{CMBPAST}. In figure \ref{f8} 
the bubbly distribution are taken from the $x=50\%$ 
case shown in figure \ref{f7}. As we anticipated, 
and as it is evident looking at figures \ref{f6},\ref{f7}, 
the power from each one of the two perturbations mechanism 
approximatively adds. In figure \ref{f8} as example of the 
effect of varying $\delta$ is shown: we plotted a 
case in which the volume fraction is $x=10\%$ but 
$\delta =2\cdot 10^{-2}$. 
As a final comment, we wish to point out how there 
are cases in which the bubbles are present, filling 
a significant fraction of the volume and having 
properties able to influence the structure formation 
process between decoupling and the present, but that 
produce CMB power that is subdominant with respect 
to the Gaussian CDM one: this happens roughly 
for $C^{T}\le 10^{-10}$, equivalent to 
$C^{P}\le 10^{-12}$. This is not surprising, 
since in these kind of non-Gaussian models, the 
power spectrum alone does not fix completely 
the underlying perturbations properties, since, 
as it is evident from its definition (\ref{cl}), it 
is sensitive on the anisotropy power at a given 
angular scale. 

\section{Conclusion}
\label{conclusion}

The most known inflationary model leaves traces 
in the form of Gaussian perturbations.
With these hypothesis, it uniquely marks the 
angular power spectrum of the cosmic microwave background (CMB) 
anisotropies. However high energy physics may be 
more complicated and may leave other (and richer) traces, 
in the form of non-Gaussian perturbations. 
During inflation, quantum tunneling phenomena may have
occurred, and the nucleated bubbles may have been stretched
to large and observable scales by the inflationary expansion
itself. Here we predict the imprint of bubbles on the
CMB, both on the polarization and temperature 
anisotropies. 
We consider an analytical bubbly density perturbation
at the beginning of the radiation era, parameterized by 
its comoving radius $R$ and density contrast $\delta$. We 
perform its evolution via the linear cosmological 
perturbation theory. We find that it uniquely marks both the 
polarization and temperature anisotropies. Then we consider 
a parametric distribution of bubbles and we predict its 
imprint on the CMB angular power spectra, as a function 
of the bubbles abundance and properties. 

During the evolution toward decoupling, we take some photos 
of the bubbly CMB perturbation at different times. As expected,
we find that the initial condition remains unchanged until
the horizon reenter occurs, and thereafter the perturbation starts
to oscillate. This behavior generates sound waves propagating
outward with the sound velocity of the photon-baryon fluid
and reaching the sound horizon at the time we are examining
the system. At decoupling, we get the CMB anisotropies by 
considering the bubbly perturbation intersecting the last
scattering surface. The dependence of the signal on the 
position of the bubble along the line of sight has been 
considered. 

The CMB polarization image of an
inflationary bubble at decoupling 
is a series of rings concentric with the
center of the bubble, extending on an angular size 
corresponding to the sound horizon at decoupling 
($\simeq 1^{o}$ in the sky); each ring is characterized by its 
own value of the Stokes parameters, giving the difference 
of the temperature fluctuations projected on the axes 
parallel and perpendicular to the plane formed by 
the radial and photons propagation directions. 
As an important difference with respect to the case
of temperature anisotropy, that reflects also in the 
CMB power spectrum from a distribution of bubbles, 
the light coming from the center of the bubble is not polarized, 
since the radial propagation in spherical symmetry is an
axial symmetric problem, so that no preferred axis exists on the 
polarization plane. Instead, temperature anisotropies are 
generally characterized by a strong central spot, together to the 
external rings. This is an important difference, that we prepare
to investigate for what concerns the impact on the CMB 
angular power spectrum. Finally, the amplitude of the CMB 
fluctuation follows the known estimates for linear perturbations 
with size $R$ and density
contrast $\delta$: $\delta T/T\simeq\delta\cdot (R/H^{-1})^{2}$
where $\delta$ and the Hubble radius are computed at decoupling; 
the amplitude of the polarization is roughly 10$\%$ 
of the pure temperature one, as the ordinary Gaussian
perturbations. 

To compute the CMB power spectra induced by a population 
of bubbles, we simulate $10^{o}\times 10^{o}$ CMB sky 
patches in which they are placed randomly and fill a 
fraction $x$ of the volume. 
By taking the angular power spectrum, we analyze first the impact 
of the bubbles only on polarization and temperature. 
We find the signal lying entirely on sub-degree 
angular scales, characterized by a series of narrow 
peaks for polarization and a broader one for temperature, 
reflecting the signals from each bubble singularly. 
The amplitudes of these peaks depends in a rather 
simple manner by volume fraction, dimension and 
central density contrast of the bubbles, from which 
we infer an empirical formula giving the approximate 
bubbly CMB power; we also use this formula to constrain 
the bubbly distribution parameters starting from 
existing sub-degree CMB observations. 
When considering bubbles together with Gaussian 
fluctuations from slow roll inflation, as realistical 
if a bubble nucleating era occurred before the 
end of inflation, we find Gaussian and bubbly 
CMB power spectra approximatively add because of their 
uncorrelation. 

Bubbly density perturbations are remnant of inflationary 
tunneling phenomena. Their detection would bring an invaluable 
insight into the hidden sector of high energy physics. In 
this work CMB bubbly signals have been computed and could be 
compared with the data from the future high resolution 
observations provided by the MAP and the Planck 
missions in the next decade. 

\vskip 1.in

The first part of this work was written at the NASA/Fermilab 
Astrophysics center. C.B. warmly thank the hospitality of the 
Theoretical Astrophysics Group. We are also grateful 
to Luca Amendola and Franco Occhionero for useful discussions.

We thank Eric Hivon and Krzysztof M. Gorski for  
their Healpix sphere pixelization that we extensively 
used in this work. We wish also to thank Davide Maino for his 
help in producing CMB maps.

\newpage


\end{document}